\documentstyle[epsfig,amssym]{europhys}
\newif\ifboo \boofalse

\begin{document}

\newcommand{\1}{{\bf \scriptstyle 1}\!\!{1}}
\newcommand{\I}{{\rm i}}
\newcommand{\p}{\partial}
\newcommand{\D}{^{\dagger}}
\newcommand{\bx}{{\bf x}}
\newcommand{\bq}{{\bf q}}
\newcommand{\bk}{{\bf k}}
\newcommand{\bv}{{\bf v}}
\newcommand{\bp}{{\bf p}}
\newcommand{\bu}{{\bf u}}
\newcommand{\bA}{{\bf A}}
\newcommand{\bB}{{\bf B}}
\newcommand{\bK}{{\bf K}}
\newcommand{\bL}{{\bf L}}
\newcommand{\bP}{{\bf P}}
\newcommand{\bQ}{{\bf Q}}
\newcommand{\bS}{{\bf S}}
\newcommand{\bH}{{\bf H}}
\newcommand{\balpha}{\mbox{\boldmath $\alpha$}}
\newcommand{\bsigma}{\mbox{\boldmath $\sigma$}}
\newcommand{\bSigma}{\mbox{\boldmath $\Sigma$}}
\newcommand{\bomega}{\mbox{\boldmath $\omega$}}
\newcommand{\bpi}{\mbox{\boldmath $\pi$}}
\newcommand{\bphi}{\mbox{\boldmath $\phi$}}
\newcommand{\bnabla}{\mbox{\boldmath $\nabla$}}
\newcommand{\bmu}{\mbox{\boldmath $\mu$}}
\newcommand{\bepsilon}{\mbox{\boldmath $\epsilon$}}

\newcommand{\iLambda}{{\it \Lambda}}
\newcommand{\cL}{{\cal L}}
\newcommand{\cH}{{\cal H}}
\newcommand{\cU}{{\cal U}}
\newcommand{\cT}{{\cal T}}

\newcommand{\be}{\begin{equation}}
\newcommand{\ee}{\end{equation}}
\newcommand{\bea}{\begin{eqnarray}}
\newcommand{\eea}{\end{eqnarray}}
\newcommand{\beqa}{\begin{eqnarray*}}
\newcommand{\eeqa}{\end{eqnarray*}}
\newcommand{\nn}{\nonumber}
\newcommand{\DD}{\displaystyle}

\newcommand{\ba}{\left[\begin{array}{c}}
\newcommand{\baa}{\left[\begin{array}{cc}}
\newcommand{\baaa}{\left[\begin{array}{ccc}}
\newcommand{\baaaa}{\left[\begin{array}{cccc}}
\newcommand{\ea}{\end{array}\right]}

\euro{}{}{}{}
\Date{}
\shorttitle{}

\title{Reply to the comment of Chudnovsky\&Garanin on "Spin relaxation in Mn$_{12}$-acetate"}

\author{Michael N.~Leuenberger\cite{email1} and Daniel Loss\cite{email2}}
\institute{Department of Physics and Astronomy, University of Basel \\
Klingelbergstrasse 82, 4056 Basel, Switzerland}

\rec{}{}

\pacs{
\Pacs{75}{45.+j}{Macroscopic quantum phenomena in magnetic systems}
\Pacs{75}{50.Xx}{Molecular Magnets}
\Pacs{75}{30.Pd}{Surface magnetism}
}

\maketitle

Angular momentum conservation is a physical law that must be obeyed in
any closed system.
However, it is well known from standard textbooks that in the case of
interactions with phonons a part of the total momentum and of the total
angular momentum can be absorbed by the whole crystal and is then no
longer considered (cf. e.g. the M\"ossbauer effect\cite{Yosida} and the
Umklapp process\cite{Ashcroft}). Therefore, there is no such general
reason why the spin-phonon couplings 
\be
g_1(\epsilon_{xx}-\epsilon_{yy})\otimes(S_x^2-S_y^2)+
\frac{1}{2}g_2\epsilon_{xy}\otimes\{S_x,S_y\}
\label{sp}
\ee
must vanish in the spin-phonon Hamiltonian ${\cal H}_{\rm sp}$ (see
Eq.~(6) in Ref.~\cite{LLPRB}) due to angular momentum conservation as
incorrectly claimed in Ref.~\cite{CG}. Both terms in Eq.~(\ref{sp}) were
used e.g. in Refs.~\cite{Hartmann}, \cite{Fort}, \cite{LLEPL}, and
\cite{LLPRB} to describe phonon-assisted tunneling in Mn$_{12}$.
Furthermore, such terms were derived in Refs.~\cite{Vleck} and
\cite{Mattuck} many years ago by means of perturbation theory: These
authors combined two spin-orbit interactions $\cH_{\rm
SO}=\lambda\bL\cdot\bS$ with one orbit-lattice interaction $\cH_{\rm
OL}$, which yields a third-order term of the form\cite{Vleck}
\be
\sum_{j,k}\frac{\cH_{ij}\cH_{jk}\cH_{ki'}}{\varepsilon_{ij}\varepsilon_{ik}},
\label{perturbation}
\ee
where $\varepsilon_{ij}=\varepsilon_i-\varepsilon_j$ are differences
between the eigenenergies of the unperturbed Hamiltonian $\cH_{\rm
O}+\cH_{\rm L}$. $\cH_{\rm O}$ is the Hamiltonian of the free ion, and
$\cH_{\rm L}$ the Hamiltonian of the lattice. Note that the
orbit-lattice interaction
\be
\cH_{\rm OL}=\sum_f\frac{\p V}{\p
Q_f}Q_f+\frac{1}{2}\sum_{ff'}\frac{\p^2 V}{\p Q_f\p Q_{f'}}
Q_fQ_{f'}+\cdots,
\label{OL}
\ee
where $V$ is the energy due to the crystalline electric field,
includes terms that are {\it linear in normal displacements $Q_f$},
which arise from the Jahn-Teller effect\cite{Vleck,Vleck2,Bersuker}. As
a matter of fact, the Jahn-Teller effect is very strong in
Mn$_{12}$\cite{Lis}. In order to provide the connection between
Eq.~(\ref{OL}) and Eq.~(\ref{sp}), we note that the strain components
$\epsilon_{\alpha\beta}$, $\alpha,\beta=x,y,z$, are proportional to the
Fourier transforms $u_\bq$ of the displacements
$Q_f$\cite{Hartmann,LLPRB}, which can be represented by phonon creation
and annihilation operators with wave vectors $\bq$, i.e. $u_\bq\propto
(a_\bq+a_\bq\D)$\cite{Hartmann,LLPRB,Mattuck}. Obviously,
Eqs.~(\ref{perturbation}) and (\ref{OL}) imply Eq.~(\ref{sp}) and that
part of the angular momentum is absorbed by the whole lattice, which is
as usual assumed to be infinitely heavy and of infinite moment of
inertia.

Moreover, it was shown generally in Ref.~\cite{Mattuck} that the
spin-phonon coupling constants corresponding to our $g_i$, $i=1,2,3,4$,
are all of the same order of magnitude. This result was also obtained by
Van Vleck\cite{Vleck} for Cr$^{3+}$.

We agree with Ref.~\cite{CG} that the $g_1$ spin-phonon coupling term
given in Eq.~(3) of Ref.~\cite{LLEPL} and in Eq.~(6) of
Ref.~\cite{LLPRB} is not due to rigid rotations only, as first pointed
out to us by J.~Villain. However, we know that
$g_4=2A$\cite{Hartmann,LLEPL,LLPRB,Melcher}, from which we obtain that
$g_1\approx A$\cite{Vleck,Mattuck}. Thus, terms of the form (\ref{sp})
do exist in Mn$_{12}$ in contrast to the incorrect claim made in
Ref.~\cite{CG}. 

Finally we conclude with the following erratum to Refs.~\cite{LLEPL} and
\cite{LLPRB}: The equation $g_1=A$ (see Ref.~\cite{LLEPL} and Eq.~(19)
in Ref.~\cite{LLPRB}) must be changed to
\be
g_1\approx A.
\label{g_1}
\ee
Our fits shown in Refs.~\cite{LLEPL} and \cite{LLPRB}, which are in good
agreement with experimental data, confirm Eq.~(\ref{g_1}).
We emphasize that all our conclusions presented in Refs.~\cite{LLEPL}
and \cite{LLPRB} remain unaffected by this change.

%*************************************************************************

\end{document}